\documentclass[aps,prd,superscriptaddress,nofootinbib,amsmath,amsfonts,preprintnumbers,groupedaddress,10pt,english]{revtex4-1}
\usepackage{amsmath}
\usepackage{amssymb}
\usepackage{babel}
\usepackage{wrapfig}
\usepackage{cancel}

\usepackage{relsize,exscale}
\makeatletter

\usepackage{array,multirow,graphicx}
\usepackage{dcolumn}
\usepackage{newlfont}
\usepackage{bm}
\usepackage[colorlinks,citecolor=blue,urlcolor=blue,linkcolor=blue]{hyperref}
\usepackage[figtopcap]{subfigure}
\usepackage{color}
\usepackage{verbatim}

\usepackage{scalerel}
\usepackage{tikz}
\usetikzlibrary{svg.path}
\definecolor{orcidlogocol}{HTML}{A6CE39}
\graphicspath{{./}{Figs/}}
\begin{document}
\tolerance=5000

\title{Spherically symmetric solution in quadratic non-metricity gravity}
\author{G.~G.~L.~Nashed
}

\email{nashed@bue.edu.eg}
\affiliation {Centre for Theoretical Physics, The British University in Egypt, P.O. Box
43, El Sherouk City, Cairo 11837, Egypt}
\affiliation {Center for Space Research, North-West University, Potchefstroom 2520, South Africa}

\author{Kazuharu Bamba
}

\email{bamba@sss.fukushima-u.ac.jp}
\affiliation {Faculty of Symbiotic Systems Science, Fukushima University, Fukushima 960-1296, Japan.}


\begin{abstract}
We explicitly find an exact spherically symmetric solution in quadratic non-metricity gravity. We show that the quadratic term acts as a cosmological constant. This solution contradicts all the claims in the literature that there is no spherically symmetric solution for higher-order non-metricity gravity. Moreover, we demonstrate that for the charged field equations, the solution can be identical to the non-charged case. This is because the off-diagonal components of the field equation do not feel the effect of the charge.
\end{abstract}



\maketitle

\section{Introduction}\label{S1}
Early in 1900s, Einstein presented his general relativity (GR) theory, transforming our idea of the universe. Supported by numerous precise observations, GR has uncovered various hidden aspects of the universe, fueling significant progress in modern cosmology. Supernova observations, for instance, have demonstrated that the universe is presently experiencing a period of accelerated expansion \cite{SupernovaCosmologyProject:1998vns}. Strong evidence suggests that mysterious forces, such as dark matter (DM) and dark energy (DE), significantly influence the structure of the universe. Unraveling the nature of this elusive energy remains a core challenge in current research. Within GR, the cosmological constant, $\Lambda$, provides a simple approach to modeling vacuum energy \cite{Sahni:1999gb}. Yet, such scenario  presents challenges such as fine-tuning and the coincidence problem \cite{Yang:2009ae, Velten:2014nra}. The above challenges imply that GR might not be the ideal framework for explaining gravity on cosmic scales. Although advancements in understanding cosmic acceleration have been modest, research into modified gravity theories (MTG) remains vital, providing strong alternatives to GR that may help resolve current issues. In the last two decades, numerous studies on MTGs have sought to enhance our comprehension of the universe's framework \cite{Wang:2019ufm, Mandal:2020rfm, Wang:2018duq, Arora:2020lsr}.

General relativity explains the interactions of gravitational field through the connection of Levi-Civita using the Riemannian spacetime, positing a geometry that is devoid of torsion and non-metricity. Nonetheless, a general affine connection can be articulated in a wider context~\cite{Hehl:1976kj, Hehl:1994ue}. Teleparallel gravity, which serves as an alternative to GR, defines gravitational interactions via torsion $\mathbb{T}$~\cite{Aldrovandi:2013wha} and employs the Weitzenb\"ock connection in its teleparallel version of GR, indicating the absence of curvature and non-metricity~\cite{Maluf:2013gaa}.
Ong \textit{et al.}~\cite{Ong:2013qja} have found that a generic $f(\mathbb{T})$ theory is likely to encounter specific issues, such as the existence of superluminal propagating modes \cite{ElHanafy:2014efn}. The presence of these modes can be identified through the characteristic equations that describe the dynamics in $f(\mathbb{T})$ gravity and/or by analyzing the Hamiltonian structure of the theory using Dirac constraint analysis. Hu \textit{et al.}~\cite{Hu:2022anq} investigated the formulation of $f(\mathbb{Q})$ gravity, where the gravitational action is constructed using the non-metricity tensor $\mathbb{Q}_{\mu \nu \rho}$. The associated Lagrangian density is expressed as a function of this quantity, allowing for deviations from standard general relativity by adopting a general function $f(\mathbb{Q})$. To study the theory from a Hamiltonian perspective, it is necessary to rewrite it using canonical variables. Using the Dirac--Bergmann algorithm, D'Ambrosio \textit{et al.}~\cite{DAmbrosio:2023asf} carried out a comprehensive Hamiltonian investigation of $f(\mathbb{Q})$ gravity and identified the count of independent dynamical degrees of freedom present in the theory.

In a related development, Hu \textit{et al.}~\cite{Hu:2023gui} identified the presence of a scalar mode in the theory, which was shown to carry a negative kinetic term indicating ghost-like behavior. Notably, this result is independent of the choice of gauge, including the coincident gauge often employed in both the symmetric teleparallel equivalent of general relativity and in $f(\mathbb{Q})$ models. To gain further insight into the behavior of the scalar mode, the authors \cite{Hu:2023gui} re-expressed the four scalar fields originally introduced as St\"uckelberg fields. 
 The inherent presence of ghosts within the symmetric teleparallel framework has been explained. Gomes \textit{et al.}~\cite{Gomes:2023tur} advanced the theoretical framework of $f(\mathbb{Q})$ gravity by studying perturbative behavior across three different spatially flat cosmological scenarios.

In the context of black holes, D'Ambrosio \textit{et al.}~\cite{DAmbrosio:2021zpm} have systematically studied the field equations of $f(\mathbb{Q})$ gravity for spherically symmetric and stationary metric-affine spacetimes. These spacetimes are characterized by a metric along with a flat, torsionless affine connection. In the symmetric teleparallel equivalent of general relativity, the connection is pure gauge and thus unphysical. However, in the non-linear extension $f(\mathbb{Q})$, the connection is promoted to a dynamical field, which alters the physical behavior of the system.
In a cosmological framework based on Weyl-Cartan spacetime, the Weitzenb\"ock connection posits that the total curvature and scalar torsion equal zero \cite{Haghani:2012bt}. Riemann-Cartan spacetime resembles the teleparallel version of GR in the absence of non-metricity. Symmetric teleparallel gravity represents another method that presumes zero curvature and torsion, while also including non-metricity $\mathbb{Q}$ \cite{Nester:1998mp}. Expanding these ideas results in $f(\mathbb{Q})$ gravity \cite{BeltranJimenez:2017tkd}, which has attracted scholarly attention \cite{Conroy:2017yln,Jarv:2018bgs,Nashed:2024ush,Latorre:2017uve,Nashed:2025zpw,Harko:2018gxr,Nashed:2025qeu,Nashed:2025usa,Hohmann:2018wxu} by presenting fresh viewpoints on geometry, gravity, and cosmic events.

In the framework of metric-affine geometry, distinct geometric models arise depending on which fundamental quantities are constrained to vanish. For example, when non-metricity is set to zero ($\mathbb{Q}=0$), the geometry reduces to Riemann-Cartan. Eliminating curvature ($\mathbb{R}=0$) gives rise to the teleparallel formulation, whereas the absence of torsion ($\mathbb{T}=0$) leads to purely non-metric geometries. Further specialization occurs when two of these geometric quantities vanish simultaneously: the condition $\mathbb{R} = \mathbb{Q} = 0$ defines the Weitzenb\"ock geometry (also referred to as teleparallelism); the case $\mathbb{T} = \mathbb{Q} = 0$ yields the Riemannian limit; and $\mathbb{R} = \mathbb{T} = 0$ corresponds to symmetric teleparallel geometry. If all three quantities are simultaneously zero, the manifold becomes equivalent to flat Minkowski spacetime.

It is important to highlight that general relativity (GR) can be expressed through various equivalent or extended formulations. One notable example is the \emph{teleparallel equivalent of general relativity} (TEGR) \cite{Nashed:2001im,Nashed:2019yto,Nashed:2014sea,Dialektopoulos:2019mtr,Barros:2020bgg,Nashed:2015pga,BeltranJimenez:2019tme,Nashed:2005kn,Bajardi:2020fxh,Nashed:2001cp,Ayuso:2020dcu,Flathmann:2020zyj,Khyllep:2021pcu,DAmbrosio:2020nev}, which formulates gravity in a spacetime where both curvature and non-metricity vanish, leaving torsion as the main dynamical entity. Another comparable representation is the \emph{symmetric teleparallel equivalent of general relativity} (STEGR) \cite{Adak:2004uh,Adak:2005cd,Adak:2008gd,Mol:2014ooa,BeltranJimenez:2018vdo,Gakis:2019rdd}, characterized by the absence of both curvature and torsion, with non-metricity playing the central role. In these equivalent models, the action is constructed from a scalar quantity derived from non-metricity, denoted by $\mathbb{Q}$.

Extensive reviews and comparative studies of these different frameworks are available in Refs.~\cite{Capozziello:2022zzh,Heisenberg:2018vsk}.

This work undertakes a comparative examination of three commonly studied extensions of GR: $f(\mathbb{R})$, $f(\mathbb{T})$, and $f(\mathbb{Q})$ gravity, with a particular focus on their implications for cosmic inflation. Both the potential slow-roll (PSR) and Hubble slow-roll (HSR) techniques are employed to analyze inflationary dynamics in each of these models \cite{Capozziello:2024lsz}. In particular, the viability of slow-roll inflation in the $f(\mathbb{Q})$ scenario has been thoroughly investigated in \cite{Capozziello:2022tvv}.


\section{Overview of the $f(\mathbb{Q})$ Framework}

In differential geometry, a connection plays a central role in defining the covariant derivative, enabling the comparison of tensor fields at different points on a manifold. Within Riemannian geometry, the connection is uniquely determined by the Christoffel symbols, also known as the Levi-Civita connection in the context of general relativity. This connection is both symmetric and metric-compatible.

However, when the condition of metric compatibility is relaxed, the connection becomes more general, allowing the presence of an antisymmetric component. This broader structure leads to richer geometric formulations beyond Riemannian geometry.

An important example of such a generalized connection is the affine connection, which is characterized by:

\begin{equation}
{\mathit \Gamma_{\phantom{\beta}\mu\nu}^{\beta}}{\mathit =\left\{ _{\phantom{\beta}\mu\nu}^{\beta}\right\} +\gamma_{\phantom{\beta}\mu\nu}^{\beta}+L_{\phantom{\beta}\mu\nu}^{\beta}.}
\label{conexion}
\end{equation}
In this study, ${ \left\{ _{\phantom{\beta}\mu\nu}^{\beta}\right\}}$  is the Christoffel symbol that can be expressed as:
\begin{equation}
{ \left\{ _{\phantom{\beta}\mu\nu}^{\beta}\right\}}{\mathit = =\frac{1}{2}g^{\beta\alpha}\left(g_{\nu\alpha,\mu}+g_{\alpha\mu,\nu}-g_{\mu\nu,\alpha}\right)},\label{christoffel}
\end{equation}
where $``,"$ means derivative w.r.t. the coordinates that uses in the metric.
The contorsion tensor, denoted as ${ \gamma_{\phantom{\beta}\mu\nu}^{\beta}}$,  is the second term in Eq.~\eqref{conexion} and represents the antisymmetric component of the connection.  It is calculated based on the torsion tensor in the following manner\footnote{We will denote the symmetric part by
( \ ), for example, $\zeta_{(\mu \nu)}=\frac{1}2( \zeta_{\mu \nu}+\zeta_{\nu \mu})$
and the  antisymmetric part by the square bracket [\ ],
$\zeta_{[\mu \nu]}=\frac{1}2(\zeta_{\mu \nu}-\zeta_{\nu \mu})$ .}:  \begin{align}
{  T_{\phantom{\beta}\mu\nu}^{\beta}=2\varGamma_{\phantom{\beta}\left[\mu\nu\right]}^{\beta}=-T_{\phantom{\beta}\nu\mu}^{\beta}},\qquad
 { \gamma_{\phantom{\alpha}\mu\nu}^{\beta}=\frac{1}{2}T_{\phantom{\alpha}\mu\nu}^{\beta}+T_{(\mu\phantom{\beta}\nu)}^{\phantom{\mu}\beta}}.\label{tns_tor} 
\end{align}
The tensor ${L_{\phantom{\beta}\mu\nu}^{\beta}}$ is the last expression in Eq.~(\ref{conexion}) and is defined as:
\begin{equation}
{ L_{\phantom{\alpha}\mu\nu}^{\beta}=\frac{1}{2}\mathbb{Q}_{\phantom{\alpha}\mu\nu}^{\beta}
-\mathbb{Q}_{(\mu\phantom{\beta}\nu)}^{\phantom{\mu}\beta}=L_{\phantom{\beta}\nu\mu}^{\beta}},\label{disf}
\end{equation}
where $\mathbb{Q}_{\phantom{\alpha}\mu\nu}^{\beta}$  is defined by the non-metricity tensor and takes the form:
\begin{equation}
{ \mathbb{Q}_{\beta\mu\nu}\equiv \nabla_{\beta}g_{\mu\nu}=g_{\mu\nu,\beta}}-\Gamma_{\phantom{\beta}\mu\beta}^{\alpha}g_{\nu \alpha}-\Gamma_{\phantom{\beta}\nu\beta}^{\alpha}g_{\mu \alpha}.   \label{tns_nmetric}
\end{equation}
It is clear that there is a relationship between the affine connection Eq.~\eqref{conexion} and the covariant derivative $\nabla_\mu$. To ease the complexity of the field equations, it is useful to introduce the superpotential, which takes the following shape:

\begin{equation}
{P_{\phantom{\alpha}\mu\nu}^{\beta}=-\frac{1}{2}L_{\phantom{\alpha}\mu\nu}^{\beta}-\frac{1}{4}\left[\left(
\tilde{\mathbb{Q}}^{\beta}-\mathbb{Q}^{\beta}\right)g_{\mu\nu}+\delta_{(\mu}^{\beta}\mathbb{Q}_{\nu)}\right]},\label{superpot}
\end{equation}
where
\begin{align}
 {\mathbb{Q}_{\alpha}=g^{\mu\nu}\mathbb{Q}_{\alpha\mu\nu}=\mathbb{Q}_{\alpha\phantom{\nu }\nu}^{\phantom{\alpha}\nu}, \qquad
\tilde{\mathbb{Q}}_{\alpha}=g^{\mu\nu}\mathbb{Q}_{\mu\alpha\nu}=\mathbb{Q}_{\phantom{\nu}\alpha\nu}^{\nu}},\end{align} are  the traces of the non-metricity tensor.
\par
Thus, we obtain a simplified formula for the non-metricity scalar by contracting the non-metricity tensor \eqref{tns_nmetric} with the superpotential \eqref{superpot} as:
\begin{equation}
{    \mathbb{Q}=-\mathbb{Q}_{\beta\mu\nu}P_{\phantom{\alpha}}^{\beta\mu\nu}}.\label{scalarQ}
\end{equation}
\par
The Levi-Civita connection dictates the curvature tensor in GR as:
\begin{equation}
{R_{\phantom{\beta}\mu\alpha\nu}^{\beta}=\partial_{\alpha}\Gamma_{\phantom{\beta}\nu\mu}^{\beta}-
\partial_{\nu}\Gamma_{\phantom{\beta}\alpha\mu}^{\beta}+\Gamma_{\phantom{\beta}\alpha\rho}^{\beta}
\Gamma_{\phantom{\rho}\nu\mu}^{\rho}-\Gamma_{\phantom{\beta}\nu\rho}^{\beta}\Gamma_{\phantom{\rho}\alpha\mu}^{\rho}}.    \label{tns_Riem}
\end{equation}
The above  tensor can be reduced in the manner described below to yield the Ricci tensor:
\begin{equation}
  {  R_{\mu\nu}=R_{\phantom{\beta}\mu\beta\nu}^{\beta}}.
\end{equation}
The Ricci scalar is obtained by summing the elements of the Ricci tensor as follows:
\begin{equation}
 { R= g^{\mu\nu}R_{\mu\nu}}.
\end{equation}

The Riemann tensor \eqref{tns_Riem} can be represented in another way by analyzing it with the affine connection as:
\begin{equation}
{R_{\phantom{\beta}\alpha\mu\nu}^{\beta}=\overset{C}{R}{}_{\phantom{\beta}\alpha\mu\nu}^{\beta}+
\overset{C}\nabla_{\mu}V_{\phantom{\beta}\nu\alpha}^{\beta}-\overset{C}\nabla_{\nu}V_{\phantom{\beta}\mu\alpha}^{\beta}
+V_{\phantom{\beta}\mu\rho}^{\beta}V_{\phantom{\rho}\nu\alpha}^{\rho}-V_{\phantom{\beta}\nu\rho}^{\beta}V_{\phantom{\rho}
\mu\alpha}^{\rho}}\,.\label{Tns_Riem_Trans}
\end{equation}
In this expression, we describe $R_{\phantom{\beta}\alpha\mu\nu}^{\beta}$ using the derivative $\overset{C}\nabla$ and the affine connection $\overset{C}{R}{}_{\phantom{\beta}\alpha\mu\nu}^{\beta}$. Quantities related to the Christoffel symbol are represented by the symbols $ \overset{C}{\Gamma}{}_{\phantom{\beta}\alpha\mu}^\beta$ and the one shown in Eq.~\eqref{christoffel}, while the expression $V^\beta_{\phantom{\beta}\mu\nu}$ is a tensor that has the following definition:
\begin{equation}
  {  V^\beta_{\phantom{\beta}\mu\nu}=\gamma^\beta_{\phantom{\beta}\mu\nu}+L^\beta_{\phantom{\beta}\mu\nu}}.
\end{equation}
\par
Moreover, formula \eqref{Tns_Riem_Trans} simplifies for the proper contractions on the Riemann tensor at $T_{\phantom{\alpha}\mu\nu}^{\beta}=0$ when a connection without torsion is taken into consideration:
\begin{equation}
{R=\overset{C}{R}-\mathbb{{Q}}+\overset{C}\nabla_{\beta}\left(\mathbb{Q}^{\beta}-\tilde{\mathbb{Q}}^{\beta}\right)},\label{scalar_Ric}
\end{equation}
The Ricci scalar, denoted by $\overset{C}{R}$, is constructed using the Christoffel symbols.

\par
By imposing the teleparallel condition, where the curvature scalar $R$ vanishes, one can establish a generalized formulation that relates the Ricci scalar to the non-metricity scalar $\mathbb{Q}$. This approach yields a teleparallel geometric structure within a flat spacetime background:
\begin{equation}
\overset{C}{R} = \mathbb{Q} - \overset{C}{\nabla}_{\beta} \left( \mathbb{Q}^{\beta} - \tilde{\mathbb{Q}}^{\beta} \right).
\label{scalar_Ric2}
\end{equation}
\noindent
This expression reveals that the Ricci scalar differs from the non-metricity scalar by a total divergence term, often interpreted as a boundary contribution:
\begin{equation}
B_\mathbb{Q} = \overset{C}{\nabla}_{\beta} \left( \mathbb{Q}^{\beta} - \tilde{\mathbb{Q}}^{\beta} \right).
\label{boundary}
\end{equation}

\noindent
Accordingly, the action for non-metricity-based gravity in four spacetime dimensions takes the following form:
\begin{align}
\label{g2}
\mathcal{S}=\int d^4 x \frac{\sqrt{-g} }{2\kappa^2}\left\{f(\mathbb{Q})-\Lambda+ 2\kappa^2 {\cal L}_{ em}~
\right\}\, ,
\end{align}
 where ${\cal L}_{
em}=-\frac{1}{2}{ F}\wedge ^{\star}{F}$ represents the Maxwell field Lagrangian,
where  $F = dA$, and  $A=A_{\mu}dx^\mu$, is the 1-form of the  electromagnetic
potential \cite{Awad:2017tyz}.
The variation of Eq.~(\ref{g2}) w.r.t. $g_{\mu\nu}$ and matter fields gives \cite{Heisenberg:2023lru}:
\begin{align}
\label{GBeq}
&0= \left[\frac{2}{\sqrt{-g}}\nabla_\gamma (\sqrt{-g}P^\gamma\:_{\mu\nu}) + \frac{1}{2}g_{\mu\nu}[\mathbb{Q}-\Lambda]+(P_{\mu\gamma\beta}\mathbb{Q}_\nu\:^{\gamma\beta} - 2\mathbb{Q}_{\gamma\beta\mu}P^{\gamma\beta}\:_\nu)\right]+\frac{\kappa^2}{2}\kappa{{{\cal
T}^{{}^{{}^{^{}{\!\!\!\!\scriptstyle{em}}}}}}}_{\mu \nu} , \nonumber\\
&0=\partial_\nu \left( \sqrt{-g} F^{\mu \nu} \right)\, .
\end{align}
In this study, ${ {{{\cal T}^{{}^{{}^{^{}{\!\!\!\!\scriptstyle{em}}}}}}}^\nu_\mu}$ is the tensor representing the electromagnetic field's energy-momentum, defined as:
\begin{align}
{{{\cal
T}^{{}^{{}^{^{}{\!\!\!\!\scriptstyle{em}}}}}}}^\nu_\mu=F_{\mu \alpha}F^{\nu \alpha}-\frac{1}{4} \delta_\mu{}^\nu F_{\alpha \beta}F^{\alpha \beta}.\end{align}
Here, as is customary, $\mathcal{T}_{\mu \nu}$ denotes the tensor of the energy-momentum of matter, specifically
\begin{align}\label{EMT}
   \mathcal{T}_{\mu \nu}= -\frac{2}{\sqrt{-g}} \frac{\delta(\sqrt{-g}\mathcal{L}_m)}{\delta g^{\mu \nu}}\;\;\;\;.
\end{align}
In the next section, we are going to test Eq.~\eqref{GBeq} for non-charged and charged case.

 \section{Spherically symmetric solution}
 Here, we analyze the spacetime that possesses spherically symmetric and static as:
\begin{align}
\label{met1}
ds^2 = -S(r)dt^2 +\frac{dr^2}{S(r)}+r^2 \left[ d\theta^2 + \sin^2 \left(\theta\right) d\phi^2\right]\,,
\end{align}
where $S$  is an unknown function of the radial coordinate.
 The non-metricity of Eq. \eqref{met1} yields the form:
\begin{align}
\mathbb{Q}=-\frac{2(S(r)+rS'(r))}{r^2}\,.
\end{align}
The field equations (\ref{GBeq}) are expressed in the following forms through the use of the metric given  by Eq.~(\ref{met1}) as\footnote{Here, we assume that $f(\mathbb{Q})=\mathbb{Q}+\alpha \mathbb{Q}^2$, where $\alpha$ has a unit of $\mathbb{Q}^{-1}$.}:
\begin{align}
&\mbox{ t\,t -component:}\nonumber\\
\label{Eq2tt}
&0=-\frac{1}{{r}^{4}}\left[8\alpha S''    {r}^{2}S +\Lambda r^4 +6\alpha S'^{2}{r}^{2}+12\alpha S' rS  -{r}^{3}S'  -4\alpha S'  r-10\alpha  S^{2}-{r}^{2}S  -4 \alpha S  +{r}^{2}+q'^2r^4\right]\,, \\
&\mbox{ r\,r -component:}\nonumber\\
\label{Eq2rr}
&0= -\frac {1}{{r}^{4}}\left[-{r}^{3}S'  -{r}^{2}S +\Lambda r^4 +6\,\alpha\, S'^{2}{r}^{2}+12\,\alpha\, S' rS  +6\,\alpha\,  S ^{2}-4\,\alpha\,S  +{r}^{2}-4\, \alpha\, S' r+q'^2r^4\right]\,, \\
&\mbox{ r\,$\theta$ -component= $\theta$\, r -component:}\nonumber\\
\label{Eq2pp1}
&0=\frac{2\alpha\,\cot \theta  \left[ {r}^{2}S'' -2\,S \right] }{{r}^{5}} \,,\\
&\mbox{$\theta$ \,$\theta$ -component=$\phi$ \,$\phi$ -component:}\nonumber\\
\label{Eq2pp}
&0=-\frac {1}{2{r}^{4}}\left[-2\,{r}^{3}S' -\Lambda r^4+12\,\alpha \,S'^{2}{r}^{2}+8\, \alpha\, S'rS -12\,\alpha\,S^{2}+12\, \alpha\, S''  { r}^{2}S -{r}^{4}S'' +8\,\alpha\, S'  {r}^{3}S''-2q'^2r^4\right]\,,\\
&\mbox{and the non-vanishing of the second equation of \eqref{GBeq}:}\nonumber\\
\label{Eqm}
&0=-\frac{1}{r}[q''r+2q']\,,
\end{align}
 where $'$ symbol refers to the derivative with respect to r, as an example, $\beta'=\frac{d\beta}{dr}$. In Eq.~\eqref{Eqm}, $q\equiv q(r)$ is the electromagnetic potential. Equation \eqref{Eq2pp1} shows that either we have $\alpha=0$, which corresponds to the STEGR case, or
 \begin{align}\label{sol} {r}^{2}S'' -2\,S=0\,, \quad \mbox{ which gives} \quad  S=c_1 r^2+\frac{c_2}r\,, \end{align}
 where $c_1$ and $c_2$ are constants. If we use  Eq. (\ref{sol}) in  Eqs. (\ref{Eq2tt}),  (\ref{Eq2rr}), and (\ref{Eq2pp}) when $\Lambda=0=q$ we get:
 \begin{align}\label{cons}
 -{\frac {54\,\alpha\,{c_1}^{2}{r}^{2}-3\,c_1\,{r}^{2}-12
\,c_1\,\alpha+1}{{r}^{2}}}=0\,, \qquad -3\,c_1\, \left( 18\,c_1\,\alpha-1 \right) =0\,.\end{align}

Equation (\ref{cons}) shows that $\alpha=\frac{1}{18 c_1}$. When we substitute $\alpha=\frac{1}{18 c_1}$ into Eqs.~(\ref{Eq2tt}), (\ref{Eq2rr}), and (\ref{Eq2pp}), we can see that they are not satisfied.

Now, we assume that $\Lambda \neq 0$ and $q=0$. By substituting Eq.~\eqref{sol} into Eqs.~(\ref{Eq2tt}), (\ref{Eq2rr}), and (\ref{Eq2pp}), we obtain:
  \begin{align}\label{cons1}
 -\frac {108\,\alpha\,{c_1}^{2}{r}^{2}-6\,c_1\,{r}^{2}-24
\,c_1\,\alpha+r^2\Lambda+2}{{r}^{2}}=0\,, \qquad 108c_1{}^2\,\alpha-6c_1+\Lambda =0\,.
\end{align}
It follows from Eq.~(\ref{cons1}) that we find:
\begin{align}\label{cons2}
c_1=-\frac{1\pm\sqrt{1-12\alpha \Lambda}}{36\alpha}\,, \quad  \mbox{and} \quad \alpha=-\frac{1}{4\Lambda}\,.
\end{align}
If we substitute Eq.~(\ref{cons2}) into Eqs.~(\ref{Eq2tt}), (\ref{Eq2rr}), and (\ref{Eq2pp}), we can show that these equations are satisfied.

Now, we are going to study the case when $\Lambda \neq 0\neq q$. In this case, if we substitute Eq.~(\ref{sol}) into Eqs.~(\ref{Eq2tt}), (\ref{Eq2rr}), and (\ref{Eq2pp}), we acquire:
\begin{align}\label{eq32}
0=-{\frac {-3\,c_1{r}^{2}+54\,\alpha\,{c_1}^{2}{r}^{2}+{\Lambda}\,{r}^{2
}-12\,\alpha\,c_1+1+{r}^{2} q'^{2}}{{r}^{2}}}.
\end{align}
The substitution of $\alpha=\frac{1}{12c_1}$ into Eq.~(\ref{eq32}) yields
\begin{align}\label{ch}
 0=-\frac{3}2\,c_1+q'^{2}-{\Lambda}\,.
\end{align}
The solution of the above equation gives:
\begin{align}\label{ch1}
q(r)=\pm\frac{\sqrt{6c_1+4\Lambda}r}{2}\,.
\end{align}
If we combine Eq.~\eqref{ch1} and Eq.~\eqref{Eqm}, we can demonstrate that the solution given by Eq.~\eqref{ch1} does not satisfy Eq.~\eqref{Eqm}.

\section{Discussions}\label{S7}

In this study, we have explicitly demonstrated that there is a spherically symmetric solution within $f(\mathbb{Q})$ theory more specifically we derive a solution for the form $f(\mathbb{Q})=\mathbb{Q}+\alpha \mathbb{Q}^2$.
For this aim  we did not set the dimensional quantity $\alpha$ equal zero but we set ${r}^{2}S'' -2S\equiv \mathbb{Q}'=0$ and derive the form of $S$ which asymptotes asymptotically as AdS/dS. We have shown that the dimensional quantities related to the cosmological constant, i.e., $\alpha=-\frac{1}{4\Lambda}$. This means that the effect of the higher order of non-metricity acts as a higher cosmological constant.

Furthermore, we have found that in the charged case there is no solution that feel of the charge. This means that in frame of $f(\mathbb{Q})$ there is no charged spherically symmetric solution. The main reason for this result is Eq.~\eqref{Eq2pp1} which has no effect of the charge. This is a general results for $f(\mathbb{Q})$ theory.

An interesting and natural extension of the present work involves the investigation of charged and uncharged solutions corresponding to a more general functional form of $f(\mathbb{Q})$. The existence and physical relevance of such solutions will be explored in a forthcoming study.

\section*{Acknowledgements}
Kazuharu Bamba acknowledges the support by the JSPS KAKENHI Grant Numbers JP21K03547, 24KF0100 and Competitive Research Funds for Fukushima University Faculty (25RK011).

%

\end{document}